

\input manumac
\twelvepoint
\doublespace
\rightline{EHU-FT-92/3}
\rightline{December 1992}
\vskip 2 cm

\centerline{ \bf PRESSURE IN CHERN-SIMONS FIELD THEORY}
\bigskip
\centerline { \bf TO THREE-LOOP ORDER }

\vskip 2 cm
\centerline { M. A. Valle Basagoiti}
\vskip .5 cm
\centerline {\it Departamento de F\'\i sica Te\'orica,}
\centerline {\it Universidad del Pa\'\i s Vasco, Apartado 644, 48080 Bilbao,
Spain}
\vskip 4 cm
\centerline {\bf Abstract}
\noindent
We calculate perturbatively the pressure of a dilute gas of
anyons through second order in the anyon coupling constant, as described by
Chern-Simons field theory. Near Bose statistics, the divergences in the
perturbative expansion are exactly cancelled by a
two-body ${\delta}$-function potential which is not required near Fermi
statistics. To the order considered, we find no need for a non-hermitian
Hamiltonian.

\vfill
\eject

This letter is about a two-dimensional non relativistic system of N
particles in the thermodynamic limit, interacting with a Chern-Simons
gauge field, and interacting among themselves via a two-body
${\delta}$-function potential. This system has been shown to
obey fractional statistics [1]. From a quantum mechanical point of view,
a system of anyons can be studied by introducing a flux tube at the
position of each particle, whose magnitude ${\theta}$ will determine
the phase associated with relative particle motion. The value
${\theta = 0}$ corresponds to the anyons being bosons, whereas
${\theta = \pi}$ corresponds to fermions. In the
dilute gas regime, when the mean interparticle spacing ${n^{-1/2}}$
is much larger than the thermal wavelenght,
$\lambda_T = ({2 \pi/m T})^{1/2}$, the first correction
to the pressure comes from the reduced two-body hamiltonian. This can be
exactly solved in terms of the Aharonov-B\"ohm phase shifts [2]. Since the
full spectrum of the N-anyon system is not known, a computation of the
pressure based on N-cluster expansion [3] is not possible.

Nevertheless, full knowledge of the spectrum is not neccesary when
the anyon model is considered from the point of view of quantum
field theory. One needs the thermodynamical potential
${\Omega(\mu,T,V) = -P V}$ as a power series in the fugacity
${z=\exp (\mu /T)}$, which can be computed via perturbation theory.
This procedure is reliable when the field-theoretical model
accounts for the fractional statistics of anyons. Thus, the
question one would like to ask is this: which field-theoretical model can be
used to reliably compute higher virial coefficients (perturbatively in
${\theta}$)?

This question has been considered in a
series of papers [4-6], where a two-body non-hermitian interaction is
introduced in order to handle the divergences due to the statistical gauge
field. As a consequence, the perturbative treatment must be carefully
done, using either a harmonic regulator, or considering the system in a
box. The source of the non-hermitian interaction arises as follows: In the
regular gauge where the statistics is standard, the anyonic gas is described
by a Hamiltonian where three-body interactions of order $\theta^2$ are
present. These lead to divergences in the perturbative expansion. However, the
three-boby interactions can be removed from the Hamiltonian by a complex
gauge transformation [5], leaving a new two-body Hamiltonian with a
non-hermitian term of order $|\theta|$. Then the gauge transformation might be
understood as  performing a self-adjoint extension [9] on the original
Hamiltonian, representing an additional interaction at order $\theta$,
namely a two-body $\delta$-function.

Another viewpoint is afforded if one consider the anyonic gas as described
by a Chern-Simons gauge field coupled to bosonic or fermionic matter. In
this case, the singular three-body interactions are kept and they correspond
to the last vertex in Table 1. Even so a perturbative expansion in
$\theta$ will be reliable if divergences cancel out due a two-body
$\delta$-function. It is not {\it a priori} clear whether both approaches must
lead to the same results.

The purpose of this letter is to calculate the
contributions to the pressure to order ${\theta^2}$ in Chern-Simons field
theory. This involves a set of  three-loop diagrams, whose computation to
second order in the fugacity is our main new result. Contributions of order
$e^{3 \mu /T}$ are neglected and will be reported in future work. An
obvious prerequisite for the validity of such computation is the reproduction
of the second virial coefficient to order ${\theta^2}$. We show that
Chern-Simons field theory with the inclusion of a repulsive ${\delta}$-function
potential among particles is a model which exactly reproduces the correct
pressure when the fidutial statistics is bosonic. The strenght of this
interaction must be adjusted to cancel UV divergences. No additional
interaction is required when the fidutial statistics is fermionic. Therefore,
regardless of whether one works with the non-hermitian Hamiltonian of refs.
[4-6] or with Chern-Simons field theory, the result for the second virial
coefficient is the same.

We shall consider a spinless non relativistic self-coupled
field interacting with a Chern-Simons gauge field, with Lagrangian density
given by $$ {\cal L}(t,{\bf x}) = -{1 \over 2 \kappa}\,\partial_t\,{\bf a}
\times{\bf a} \,+ {1\over\kappa}\, a_0 B \, -
{1\over 2 \rho}\,(\nabla\cdot{\bf a})^2 \, +
\psi^\dagger i D_0 \psi - {1\over 2 m}\mid\! {\bf D}
\psi\!\mid^2 \, +\, \mu\,\psi^\dagger \psi -
{\alpha\over 4}\, (\psi^\dagger \psi)^2, $$
$$ D_0 = \partial_t + i\, a_0, \qquad \
{\bf D}=\nabla - i\,{\bf a} \eqno(1) $$
Here $B=\nabla\times {\bf a}$ is the magnetic field, $\mu$ is the chemical
potential and $\rho$ is a gauge fixing parameter. The Coulomb gauge to be used
refers to the choice $\rho=0$. Finally, $\alpha = O(\kappa)$  is the strength
of the two-body contact interaction of the form  $$V({\bf x}_1-{\bf
x}_2)={\alpha\over2}\, \delta({\bf x}_1- {\bf x}_2)\,.\eqno(2)$$
When the starting statistics is fermionic we will set $\alpha=0$.

In the
imaginary-time formalism of thermal field theory [7], the functional integral
expression for the grand partition function involves an integration over
imaginary time from $0$ to $\beta = T^{-1}$ ,
$$ Z(\mu,T,V) =
\int_{\rm (anti)periodic} {\cal D}a_0\,{\cal D}{\bf a}\, {\cal
D}\psi^\dagger{\cal D}\psi\, \exp\,\Bigl( \int_0^\beta d\tau \int
d^2\!x\, {\cal L}(t\!=\!-i\tau,{\bf x}) \Bigr)\,. \eqno(3) $$
Here (anti)periodic means that the integration over fields
is constrained so that $\psi({\bf x},\beta)=\pm \psi({\bf x},0)$ where
the (lower) upper sign refers to (fermions) bosons.

Now we can proceed to expand in a power series in $\kappa$, by using a set
of diagrammatic rules which follows from (1). These are listed in Table 1.
The diagrams describing the perturbative series for $\ln Z$ have the form
of connected closed loops. It should be noted that there will
be a factor of $\beta V$ left over for each graph, corresponding to the
extensivity of  $\ln Z$. This factor cancel out in expressions for the
pressure.

The non-zero graphs contributing to the pressure up to order $\kappa^2$ are
shown in fig.1. Note that all graphs of order $\kappa$ vanish due to the index
summatios associated with vertices. The same argument applies to graphs of
order $\alpha \kappa$, although at higher orders this is not necessarily so.
In order to keep the correct order of the operators according to their
$\tau$-values, one must insert a factor $e^{i\omega_{n}\eta}$ whenever a
particle line either closes on itself or is joined by the same instantaneous
interaction line. We take $\eta\rightarrow 0^{+}$ at the end of the
calculation. As a consequence, terms involving three bubbles in
$P_{\alpha}^{(2)}$ and in $P_{\kappa}^{(2)}$  are of order $e^{3\mu\beta}$.
This follows simply from power counting with the formulae
$$\eqalignno{
{1\over\beta}\sum_{n}{e^{i \omega_{n} \eta} \over i\,\omega_{n} + \mu - {\bf
q}^2/(2 m)} &=  -\zeta\, {1\over \exp{\bigl[\beta\,({\bf q}^2/(2 m) -
\mu)\,\bigr ]}  -\zeta} =  -\zeta\,n_{\zeta}({\bf q})\,, &(4)\cr
{1\over\beta}\sum_{n}{1  \over i\,\omega_{n} + \mu - {\bf q}^2/(2 m)} &=
-{1\over 2} -\zeta\,n_{\zeta}({\bf q})\,,&(5)\cr} $$
where $\zeta=1$ for
bosons and $\zeta = -1$ for fermions. Therefore, graphs (c) and (g) can be
neglected, since we are computing to order $e^{2\mu\beta}$.

We are now in position to derive the perturbative corrections
to the pressure
$$ P = {T\over V}\ln Z .\eqno(6)$$
Graph (a) gives
$$ P_{\alpha}^{(1)} = -{\alpha\over 2}
\int {d^2{\bf p}\over (2 \pi)^2}\,{d^2{\bf q}\over (2 \pi)^2}\,
n_{1}({\bf p})\, n_{1}({\bf q})
=-{\alpha\, e^{2\beta\mu}\over 2 \lambda_{T}^4} + O(e^{3\beta\mu}), \eqno(7)$$
and there are not first order correction with fermionic statistics.

The contribution from graph (b) may be written
$$P_{\alpha}^{(2)} =
   {\alpha^2\over 8\beta} \sum_{\nu_{n}}
\int {d^2{\bf q}\over (2 \pi)^2}\,
\bigl[\Pi^0{}_{00}(\nu_{n},{\bf q})\bigr ]^2 , \eqno(8) $$
where
$$\eqalignno{ \Pi^0{}_{00}(\nu_{n},{\bf q})&=
-\zeta {1\over\beta} \sum_{\omega_{1}}\int {d^2{\bf p}\over (2 \pi)^2}\,
{\cal G}^{0}(\omega_{1},{\bf p})\,
{\cal G}^{0}(\omega_{1}+\nu_{n},{\bf q} + {\bf p}) \cr
&=-\int {d^2{\bf p}\over (2 \pi)^2}\,
{ n_{\zeta}({\bf p}+{\bf q}) - n_{\zeta}({\bf p}) \over
i\,\nu_{n} - \varepsilon_{{\bf p}+{\bf q}} +
 \varepsilon_{\bf p } }\,.  &(9) \cr          } $$
represents the lowest-order density-density correlator (see fig.2).
At high temperature and low density, distribution
functions reduce to $n_{\zeta}({\bf p}) =
e^{\beta \mu} e^{-\beta\, {\bf p}^2/2 m}$. Then, keeping terms to second
order in the fugacity, eq.(8) gives
$$ P_{\alpha}^{(2)} =
{ \alpha^2\,m\,e^{2 \beta \mu}\over 8 \pi \,\lambda_{T}^4}
\int_{0}^{x^{\ast}} dx\,\Phi(x) , \eqno(10)$$
where $x^\ast = {\sqrt {\beta \over 4 m}}\, q^\ast$ is a cutoff and
$\Phi(x)$ is the plasma dispersion function [8],
$$\Phi(x) = 2 \int_0^x dt\,{e^{-t^2}\,t\over\sqrt{x^2 - t^2}} =
2 e^{-x^2} \int_0^x dt\, e^{t^2}.\eqno(11)$$
$\Phi(x)$ has the following limiting behavior
$$\eqalign{ \Phi(x) &= 2 x + O(x^3) \qquad x\ll 1, \cr
  \Phi(x) &= x^{-1} + O(x^{-3}) \quad x\gg 1\,. \cr} \eqno(12) $$
We have found a UV logarithmic divergence. This corresponds to the divergence
in the second Born approximation for the scattering amplitude with a
$\delta$-function potential [9].

Now we consider the ring contribution from graph (d) of fig. 1,
$$P_\kappa^{d} = {\kappa^2\over 2\beta} \sum_{\nu_{n}}
\int {d^2{\bf q}\over (2 \pi)^2}\,{\Pi^{0}(\nu_{n},{\bf q})\,
\Xi^{0}(\nu_{n},{\bf q})\over {\bf q}^2}, \eqno(13)$$
where $\Xi^{0}(\nu_{n},{\bf q})$ is the two-dimensional transverse component of
the lowest-order current-current correlation given by (see fig.2)
$$\eqalignno{ \Pi^0{}_{ij}(\nu_{n},{\bf q}) &=
-{\zeta\over m^2}{1\over\beta}\sum_{\omega_1} \int {d^2{\bf p}\over (2
\pi)^2}\,
\Bigl[({\bf p} + {{\bf q}\over 2})_i\,({\bf p} +{{\bf q}\over 2})_j\,
{\cal G}^{0}(\omega_{1},{\bf p})\,
{\cal G}^{0}(\omega_{1}+\nu_{n},{\bf q} + {\bf p}) \cr
&\qquad+ m\,\delta_{ij}\,e^{i\omega_1 \eta}\,{\cal G}^{0}(\omega_{1},{\bf
p})\,\Bigr]\cr &= -\Pi^0{}_{00}\,(\nu_{n},{\bf q})\,{\nu_n^2\over{\bf q}^2}\,
     {q_i\, q_j\over{\bf q}^2} +
    \Xi^{0}\,(\nu_{n},{\bf q})\,(\delta_{ij} - {q_i\, q_j\over{\bf q}^2})\,.
     &(14)\cr }  $$
In the classical limit, when
$n_{\zeta}({\bf p}) =
e^{\beta \mu} e^{-\beta\, {\bf p}^2/2 m}$,
$$\Xi^{0}(\nu_{n},{\bf q}) =
{1\over m \beta}\bigl(\,\Pi^0{}_{00}(\nu_{n},{\bf q}) -
   {\beta \,e^{\beta\mu}\over\lambda_T^2}\,\bigr) + O(e^{2\beta\mu}),
\eqno(15)$$
and eq. (13) gives
$$P_\kappa^{d} ={\kappa^2\, e^{2 \beta \mu} \over 8
\pi m \,\lambda_{T}^4} \int_{0}^{x^{\ast}} dx\,
\Biggl[{\Phi(x)\over x^2} - {2\over x}\Biggr]\, . \eqno(16)$$
Again a UV logarithmic divergence appears coming from the ring loop with
a seagull vertex.

We still have to compute the exchange graphs (e) and (f) of fig. 1. These give
$$\eqalignno{ P_{\kappa}^{e}&=
-\zeta{\kappa^2\over 4 m^2\,\beta^3}\sum_{\nu\,\,\omega_{1}\,\omega_{2}}
\int {d^2{\bf q}\over (2 \pi)^2}\,{d^2{\bf p}\over (2 \pi)^2}\,
{d^2{\bf k}\over (2 \pi)^2}\,
{\cal G}^{0}(\omega_{1},{\bf p})\,
{\cal G}^{0}(\omega_{1}+\nu,{\bf q} + {\bf p})\cr
&\qquad\times{\cal G}^{0}(\omega_{2}-\nu,{\bf k} - {\bf q})\,
{\cal G}^{0}(\omega_{2},{\bf k})\,
{ \bigl[\,({\bf p}-{\bf k}+{\bf q})\times{\bf q}\,\bigr]^2 \over
        ({\bf p}-{\bf k}+{\bf q})^2\,{\bf q}^2 }\,, &(17) \cr
 P_{\kappa}^{f}&=
-\zeta{\kappa^2\over m\,\beta^3}\sum_{{\nu}\,\,
\omega_{1}\,\omega_{2}} \int {d^2{\bf q}\over (2
\pi)^2}\,{d^2{\bf p}\over (2 \pi)^2}\, {d^2{\bf k}\over (2 \pi)^2}\,
e^{i\omega_1 \eta}\,{\cal G}^{0}(\omega_{1},{\bf p})\,
e^{i\omega_2 \eta}\,{\cal G}^{0}(\omega_{2},{\bf k}) \cr
&\qquad\times{\cal G}^{0}(\omega_{1} + \nu,{\bf p} +{\bf q})\,\,
{\,({\bf p}-{\bf k}+{\bf q}) \cdot {\bf q}
\over ({\bf p}-{\bf k}+{\bf q})^2\,{\bf q}^2 }\,. &(18) \cr } $$
A detailed evaluation of these terms to order
$e^{2\beta\mu}$  yields the divergent contribution
$$ P_{\kappa}^{e} + P_{\kappa}^{f} =
-\zeta{\kappa^2\, e^{2 \beta \mu}\over 4 \pi m \,\lambda_{T}^4}
\int_{0}^{x^{\ast}} dx\,\Phi(x)\,,\eqno(19)$$
where the divergence comes again from the graph (f) with a seagull vertex.

In the fermionic case, the total contribution from the gauge field
is finite by itself, giving
$$\eqalignno {P_{\kappa}^{(2)} &=
P_{\kappa}^{d} +P_{\kappa}^{e} + P_{\kappa}^{f} \cr
&={ \kappa^2\, e^{2 \beta \mu} \over 8
\pi m \,\lambda_{T}^4} \int_{0}^{\infty} dx\,
\Biggl[{\Phi(x)\over x^2} - {2\over x} + 2\,\Phi(x) \Biggr]  \cr
&= -{ \kappa^2\, e^{2 \beta \mu} \over 8
\pi m \,\lambda_{T}^4}\int_{0}^{\infty} dx\,{d\over dx}\Biggl[
  {\Phi(x)\over x} \Biggr] \cr
&= { \kappa^2\, e^{2 \beta \mu} \over 4
\pi m \,\lambda_{T}^4}\,. &(20) \cr}  $$
In the bosonic case, the divergences in the total contribution to the pressure
to order $\kappa^2$ cancel out only if $\alpha=\pm 2 \kappa/m$, and we
find      $$\eqalignno {P_{\kappa}^{(2)} &=
P_{\kappa}^{d} +P_{\kappa}^{e} +
P_{\kappa}^{f}+ P_{\alpha}^{(2)} \cr
&={ \kappa^2\, e^{2 \beta \mu} \over 8
\pi m \,\lambda_{T}^4} \int_{0}^{\infty} dx\,
\Biggl[{\Phi(x)\over x^2} - {2\over x} - 2\,\Phi(x) + 4\,\Phi(x)  \Biggr]  \cr
&= { \kappa^2\, e^{2 \beta \mu} \over 4
\pi m \,\lambda_{T}^4}\,, &(21) \cr}  $$
which is the same result as in fermionic case.

Finally, using the pressure for two-dimensional ideal quantum gases
$$ P^{(0)}(\mu,T) = \zeta\, T \lambda_T^{-2}\,{\rm Li}_2(\zeta e^{\mu/T}
)\,,\eqno(22) $$
where ${\rm Li}_2$ denotes the dilogarithm function, we
obtain the cluster expansion
$$\eqalignno{ {P(\mu,T)\over T}&=
{1\over \lambda_T^{2}} \sum_{l=1}^\infty b_l\,z^l =
   T^{-1} (P^{(0)}(\mu,T) + P_{\alpha}^{(1)} + P_{\kappa}^{(2)} + O(\kappa^3)\,
)\cr
 &= {z\over \lambda_T^{2}} + {z^2\over\lambda_T^{2}} \Biggl [{\zeta\over
4}\,\mp\, {(1+\zeta)\,\kappa\over 4 \pi}+
 {\kappa^{2}\over 8 \pi^2}\Biggr ] + O(z^3)\,. &(23) \cr}$$
When the fidutial statistics is bosonic this formula reproduces the correct
second virial coefficient by taking the upper sign corresponding to a
repulsive contact interaction of strength $\alpha = 2\kappa/m$ and putting
$\kappa = 2 \theta$. The other possible choice, $\alpha = -2\kappa/m$,
corresponds to the nonrelativistic model admitting classical self-dual
solutions [10]. When the fidutial statistics is fermionic, the suitable
choice is $\kappa =2 (\theta -\pi)$, where $\theta$ is the statistical
parameter in both cases.

To conclude, we have shown that standard perturbative Chern-Simons field
theory coupled to nonrelativistic matter exactly reproduces the second
virial coefficient. Through second order in $\theta$, all divergences cancel,
and there is no need for renormalization or resummation. A non-hermitian
Hamiltonian is not required to account for the statistical interaction between
anyons.

\vskip 1.5 cm
I thank M.A. Go\~ni and J.L. Ma\~nes for helpful comments. Financial support
under contract AEN90-0330 and UPV172.310-E035/90 is acknowledged.

\vfill
\eject
\beginsection References

\frenchspacing
\item{[1]} For a recent review of this subject see F. Wilczek, Fractional
statistics and anyon superconductivity (World Scientific, Singapore, 1990);
S. Forte, Rev. Mod. Phys. {\bf64} (1992) 193.

\item{[2]} D. P. Arovas, R. Schrieffer, F. Wilczek and A. Zee, Nucl.
Phys. B {\bf 251} (1985) 117; see also D. P. Arovas, in: Geometric Phases
in Physics, eds. A. Shapere and F. Wilczek (World Scientific, Singapore,1989)
p.284.

\item{[3]} K. Huang, Statistical mechanics (Wiley, New York, 1987) ch.10.

\item{[4]} J. McCabe and S. Ouvry, Phys. Lett. B {\bf260} (1991) 113.

\item{[5]} A. Comtet, J. McCabe and S. Ouvry, Phys. Lett. B {\bf260} (1991)
372.

\item{[6]} A. Dasni\`eres and S. Ouvry, Phys. Lett. B {\bf291} (1992) 130.

\item{[7]} J. I. Kapusta, Finite-Temperature Field Theory (Cambridge
University Press, Cambridge, 1990).

\item{[8]} A. L. Fetter and J. D. Walecka, Quantum Theory of Many-Particle
Systems (McGraw-Hill, New York, 1971), sec. 33.

\item{[9]} R. Jackiw, Delta-function potentials in two
 and three-dimensional quantum mechanics, MIT preprint CPT 1937
(January 1991); C. Manuel and R. Tarrach, Phys. Lett. B {\bf268} (1991) 222.

\item{[10]} R. Jackiw and S. Y. Pi, Phys. Rev. D {\bf42} (1990)
3500.

\vfill
\eject

\noindent
\centerline {Table 1. {\it Bare propagators and vertices}}
\vskip 0.2 cm
\hrule
\vskip 0.1 cm
\hrule

\vskip 1.3 cm
\halign{$\qquad\qquad\qquad\qquad\qquad\qquad\qquad\qquad\qquad$#\hfill\cr
${\cal G}^{0}(\omega_{n},{\bf p}) ={\displaystyle{
{1\over i\, \omega_{n} + \mu - {\bf p}^2/(2 m)}}}$\cr
\noalign{\vskip 0.1 cm}
\qquad $\omega_{n} = {\displaystyle{{2 n \pi\over
\beta}}}$ {\quad} in boson propagator \cr
\noalign{\vskip 0.1 cm}
\qquad $\omega_{n} = {\displaystyle{{(2 n + 1) \pi\over \beta}}}$ {\quad} in
fermion propagator\cr
\noalign{\vskip 1.5 cm}
${\cal D}_{0j}({\bf q}) = {\displaystyle{{i\,\kappa
\,\epsilon_{j m}\, q_m\over {\bf q}^2}}}$\cr
${\cal D}_{00}({\bf q}) = {\cal D}_{ij}({\bf q}) =0${\quad} in the Coulomb
gauge, $\rho=0$ \cr
\noalign{\vskip 2.2 cm}
$\Gamma^{\alpha} = -\alpha$\cr
\noalign{\vskip 2.2 cm}
${\Gamma}_{0}= -1$\cr
\noalign{\vskip 2.2 cm}
${\Gamma}_{j}(p,k) = {\displaystyle{
{p_j + k_j\over 2 m}}}$\cr
\noalign{\vskip 2 cm}
${\Delta}_{ij}= {\displaystyle{{1\over m} \, \delta_{ij}}}$\cr
}
\vskip 1.5 cm
\hrule
\vskip 0.1 cm
\hrule

\vfill
\eject
\beginsection Figure Captions

Fig. 1. Non zero diagrams contributing to the pressure to the second order in
$\kappa$. The sign $\pm$ refers to Bose or Fermi propagators. Graphs (c) and
(g) contribute to the third virial coefficient. Combinatoric factors are
shown in the diagram.

\noindent
Fig. 2. The self-energy of the statistical gauge field at the one-loop
level in terms of the functional derivative of the (zero) pressure to lowest
order. 1PI means that only the one-particle irreducible diagrams contribute
to $\Pi^0$.

\vfill
\eject

\end


\def\doublespace{\baselineskip=20pt plus 2pt
\lineskip=3pt minus 1pt\lineskiplimit=2pt}

\def\nofirstpagenoten{\nopagenumbers\footline={\ifnum\pageno>1\tenrm
\hss\folio\hss\fi}}
\def\nofirstpagenotwelve{\nopagenumbers\footline={\ifnum\pageno>1\twelverm
\hss\folio\hss\fi}}
\def\leaderfill{\leaders\hbox to 1em{\hss.\hss}\hfill}


\parindent=20pt
\def\narrow{\advance\leftskip by 40pt \advance\rightskip by 40pt}

\def\nonarrower{\advance\leftskip by -40pt\advance\rightskip by -40pt}

\def\boxit#1{\vbox{\hrule\hbox{\vrule\kern3pt
        \vbox{\kern3pt#1\kern3pt}\kern3pt\vrule}\hrule}}

\def\gtorder{\mathrel{\raise.3ex\hbox{$>$}\mkern-14mu
             \lower0.6ex\hbox{$\sim$}}}
\def\ltorder{\mathrel{\raise.3ex\hbox{$<$}|mkern-14mu
             \lower0.6ex\hbox{\sim$}}}
\def\dalemb#1#2{{\vbox{\hrule height .#2pt
        \hbox{\vrule width.#2pt height#1pt \kern#1pt
                \vrule width.#2pt}
        \hrule height.#2pt}}}

\font\twelvett=cmtt12 \font\twelvebf=cmbx12
\font\twelverm=cmr12 \font\twelvei=cmmi12 \font\twelvess=cmss12
\font\twelvesy=cmsy10 scaled \magstep1 \font\twelvesl=cmsl12
\font\twelveex=cmex10 scaled \magstep1 \font\twelveit=cmti12
\font\tenss=cmss10
 
 \font\ninebf=cmbx9
\font\ninerm=cmr9 \font\ninei=cmmi9
\font\ninesy=cmsy9 
\font\eightrm=cmr8
\catcode`@=11
\newskip\ttglue
\newfam\ssfam

\def\twelvepoint{\def\rm{\fam0\twelverm}
\textfont0=\twelverm \scriptfont0=\ninerm \scriptscriptfont0=\sevenrm
\textfont1=\twelvei \scriptfont1=\ninei \scriptscriptfont1=\seveni
\textfont2=\twelvesy \scriptfont2=\ninesy \scriptscriptfont2=\sevensy
\textfont3=\twelveex \scriptfont3=\twelveex \scriptscriptfont3=\twelveex
\def\it{\fam\itfam\twelveit} \textfont\itfam=\twelveit
\def\sl{\fam\slfam\twelvesl} \textfont\slfam=\twelvesl
\def\bf{\fam\bffam\twelvebf} \textfont\bffam=\twelvebf
\scriptfont\bffam=\ninebf \scriptscriptfont\bffam=\sevenbf
\def\tt{\fam\ttfam\twelvett} \textfont\ttfam=\twelvett
\def\ss{\fam\ssfam\twelvess} \textfont\ssfam=\twelvess
\tt \ttglue=.5em plus .25em minus .15em
\normalbaselineskip=14pt
\abovedisplayskip=14pt plus 3pt minus 10pt
\belowdisplayskip=14pt plus 3pt minus 10pt
\abovedisplayshortskip=0pt plus 3pt
\belowdisplayshortskip=8pt plus 3pt minus 5pt
\parskip=3pt plus 1.5pt
\setbox\strutbox=\hbox{\vrule height10pt depth4pt width0pt}
\let\sc=\ninerm
\let\big=\twelvebig \normalbaselines\rm}
\def\twelvebig#1{{\hbox{$\left#1\vbox to10pt{}\right.\n@space$}}}

\def\tenpoint{\def\rm{\fam0\tenrm}
\textfont0=\tenrm \scriptfont0=\sevenrm \scriptscriptfont0=\fiverm
\textfont1=\teni \scriptfont1=\seveni \scriptscriptfont1=\fivei
\textfont2=\tensy \scriptfont2=\sevensy \scriptscriptfont2=\fivesy
\textfont3=\tenex \scriptfont3=\tenex \scriptscriptfont3=\tenex
\def\it{\fam\itfam\tenit} \textfont\itfam=\tenit
\def\sl{\fam\slfam\tensl} \textfont\slfam=\tensl
\def\bf{\fam\bffam\tenbf} \textfont\bffam=\tenbf
\scriptfont\bffam=\sevenbf \scriptscriptfont\bffam=\fivebf
\def\tt{\fam\ttfam\tentt} \textfont\ttfam=\tentt
\def\ss{\fam\ssfam\tenss} \textfont\ssfam=\tenss
\tt \ttglue=.5em plus .25em minus .15em
\normalbaselineskip=12pt
\abovedisplayskip=12pt plus 3pt minus 9pt
\belowdisplayskip=12pt plus 3pt minus 9pt
\abovedisplayshortskip=0pt plus 3pt
\belowdisplayshortskip=7pt plus 3pt minus 4pt
\parskip=0.0pt plus 1.0pt
\setbox\strutbox=\hbox{\vrule height8.5pt depth3.5pt width0pt}
\let\sc=\eightrm
\let\big=\tenbig \normalbaselines\rm}
\def\tenbig#1{{\hbox{$\left#1\vbox to8.5pt{}\right.\n@space$}}}
\let\rawfootnote=\footnote \def\footnote#1#2{{\rm\parskip=0pt\rawfootnote{#1}
{#2\hfill\vrule height 0pt depth 6pt width 0pt}}}